\documentclass[pra,10pt,twocolumn,aps]{revtex4-1}

\usepackage{amsmath}    
\usepackage{amssymb}
\usepackage{graphicx}   
\usepackage{verbatim}   
\usepackage{color}      
\usepackage{subfigure}  

\definecolor{gray(x11gray)}{rgb}{0.75, 0.75, 0.75}

\definecolor{darksienna}{rgb}{0.24, 0.08, 0.08}

\definecolor{darkblue}{rgb}{0.0, 0.13, 0.35}

\usepackage{colortbl}

\usepackage[colorlinks=true,
			urlcolor=blue,
			citecolor=blue,
            linkcolor=blue]{hyperref}   
            
\begin{document}

\title{Vacuum birefringence detection in all-optical scenarios}
\author{Stefan Ataman}
\affiliation{Extreme Light Infrastructure - Nuclear Physics (ELI-NP)\\
30 Reactorului Street,\\
077125 M\u{a}gurele, jud. Ilfov, Romania}

\date{\today}

\begin{abstract}
In this paper we propose an all-optical vacuum birefringence experiment and evaluate its feasibility for various scenarios. Many petawatt-class lasers became operational and many more are expected to enter operation in the near future, therefore unprecedented electromagnetic fields ($E_L\sim10^{14}-10^{15}$ V/m and intensities $I_L\sim10^{21}-10^{23}$ W/cm$^2$) will become available for experiments. In our proposal a petawatt-class laser disturbs the quantum vacuum and creates a delay in a counter-propagating probe laser beam. Placing this delayed beam in one arm of   a Mach-Zehnder interferometer (MZI), allows the measurement of the vacuum refraction coefficient via a phase shift. Coherent as well as squeezed light are both considered and the minimum phase sensitivity evaluated. We show that using existing technology and with some moderately optimistic assumptions, at least part of the discussed scenarios are feasible for a vacuum birefringence detection experiment.
\end{abstract}

\maketitle

\section{Introduction}
\label{sec:introduction}
It is a well known fact that Maxwell's equations \cite{Jac99} are linear, therefore two or more electromagnetic fields propagating in vacuum -- as far as classical electrodynamics is concerned -- do not influence each other, no matter how intense they are. This state of facts would dramatically change just a few years after Dirac's remarkable equation \cite{Dir28} and its prediction of anti-matter.

Euler and Kockel \cite{Eul35} where the first to give a	theoretical formulation of optical non linearities in the quantum vacuum at the lowest orders in the electromagnetic fields. Their work was extended in \cite{Hei36} where a complete theoretical study of what is now called the Heisenberg-Euler Lagrangian is given. Decades later, Schwinger \cite{Sch51}, using the ``proper-time method'' from QED (Quantum Electro-dynamics) reconfirmed the Heisenberg-Euler result.

One consequence of the Heisenberg-Euler Lagrangian \cite{Hei36} is vacuum birefringence \cite{Kle64,Bir70}. Indeed, the quantum vacuum under the influence of external electric and/or magnetic fields behaves as if it were a birefringent (or anisotropic) material medium. Initially discussed for constant background fields only \cite{Sau31,Sch51,Kle64}, vacuum birefringence is expected for an alternating background \cite{Bre70} or in an intense laser field \cite{Aff88,Bec91,Kin16}, too.


On the experimental side, following a proposal from Iacopini \& Zavattini \cite{Iac79} and with a realistic experimental method outlined in reference \cite{Bak98}, the PVLAS collaboration was started. Their device (see e. g. Fig.~2 in references \cite{Zav12} or \cite{Del16}) is based on a linearly polarized laser passing through a strong magnetic field. In order to enhance the effect, a Fabry-Perot cavity is introduced and the output laser signal is analyzed for small rotations of its initial polarization plane. Several upgrades of the experimental setup allowed the PVLAS team to report sensitivities three orders of magnitude above the QED limit \cite{Zav12} in $2012$. The latest PVLAS results \cite{Del14,Del16} place their experimental sensitivity at a factor of $50$ above the QED limit, with no vacuum birefringence signal detected. Other magnetic field based vacuum birefringence experiments include BMV (Birefringence Magn\'etique du Vide) \cite{Cad14} and Q\&{A} \cite{Mei10}. None reported a vacuum birefringence signal.


The magneto-electric birefringence in the quantum vacuum was discussed by Rikken and Rizzo \cite{Rik00,Rik03}. In a configuration where the quantum vacuum is subjected to strong electric and magnetic fields (perpendicular to each other and perpendicular to the probe photon direction) the authors found besides the expected magnetic (``Cotton-Mouton'') and electric (``Kerr'') birefringences a magneto-electric birefringence. The problem with the magneto-optical effect, as opposed to the purely magnetic case is that using a cavity actually cancels the effect. This is due to the fact the the magneto-optical term changes sign if the probe photon reverses its direction \cite{Rik00}. Therefore, experimental proposals focused on a hybrid, two cavity setup \cite{Lui04,Lui04b,Lui05}. This design, however, does not seem to be adapted to very high intensity ($\sim$ PW class) lasers, where all optical elements are rated for a very limited number of laser shots.

The advent of petawatt class laser facilities, for example Vulcan \cite{VULCAN}, Apollon \cite{Zou15}, LFEX \cite{Miy06}, ELI-NP \cite{ELI-NP,Tes13} and ELI-BL \cite{ELI-BL} (the last two being part of the European ``Extreme Light Infrastructure'' project \cite{ELI}) has boosted various proposals to probe quantum vacuum nonlinearities in high-intensity laser experiments \cite{diP12}; vacuum birefringence is one of them.

Higher frequency probe beams (X-rays, gamma rays) \cite{Ild16,Bra17} have also been considered. At the HiBef consortium at DESY, the vacuum birefringence measurement was deemed a flagship experiment \cite{Sch16}. The proposal is based on the interaction of X-rays with a high-power laser, already considered in \cite{DiP06}. At ELI-NP \cite{Hom16,ELINP2017} a vacuum birefringence experiment based on the interaction of gamma rays with strong ($\sim{1-10}$ PW) lasers was proposed \cite{ELINP2017,Nak17} and is considered as a future experiment. It relies on the proposals of Dinu et al. \cite{Din14a,Din14b}.

To our best knowledge, no terrestrial vacuum birefringence experiment has reported a positive result. The only paper reporting a vacuum birefringence signal \cite{Mig16} is based on astrophysical observations of an isolated neutron star, however it was met with criticism \cite{Cap17}.

In this paper we discuss a new, interferometer-based and all-optical experimental setup able to measure a vacuum birefringence signal. The ``pump'' beam is a strong petawatt-class laser while the ``probe'' beam is a linearly polarized (CW or pulsed) optical laser beam. In some scenarios we also add squeezed vacuum into the second input port of the interferometer. The birefringence of the vacuum causes a small phase shift $\Delta\varphi_{QED}$ in the Mach-Zehnder interferometer, an effect that should be measurable under certain circumstances.

The phase accuracy of a Mach-Zehnder interferometer is limited by the so-called  shot noise or standard quantum limit\cite{GerryKnight,MandelWolf,Dem15} (i.e. $\Delta\varphi_{SQL}\sim1/\sqrt{\langle{N}\rangle}$ where $\langle{N}\rangle$ is the average number of photons) if classical (coherent) light is used. This would make a vacuum birefringence experiment with femto- to pico-second scale pulses infeasible with realistic power estimates for the CW probe laser. A workaround could be found by employing pulsed probe beams reaching high peak powers \cite{Bar94}. In this scenario phase stability and synchronization with pump laser are issues to be solved.

Squeezed stated of light (Yurke \cite{Yue76,Yur85}, see also \cite{GerryKnight}) can lower the phase sensitivity of an interferometer. This technique has been successfully implemented by the LIGO collaboration in order to detect gravitational waves \cite{Aba11,LIGO13}. Caves \cite{Cav81} was the first to show that introducing squeezing in the unused port of an interferometer can lower its phase sensitivity below the shot-noise limit. Experimental demonstration with a Mach-Zehnder interferometer \cite{Xia87} soon followed, proving that the concept is usable in practical measurements. Over the next decades, steady improvements in both theoretical and experimental aspects brought the sensitivity of a Mach-Zehnder interferometer to the so-call Heisenberg limit \cite{Dem15} (i.e. $\Delta\varphi_{HL}\sim1/{\langle{N}\rangle}$).

Using the Cram\'er-Rao bound and Fisher information, Pezz\'e and Smerzi \cite{Pez08} showed that by injecting a coherent state in one port of an interferometer and squeezed vacuum in the other, the Heisenberg limit can be achieved if roughly half of the input power goes into squeezing. Lang and Caves \cite{Lan13, Lan14} confirmed this result using Fisher information and showed it to be optimal for the class of coherent $\otimes$ squeezed vacuum type of input states. Sparaciari, Olivares and Paris \cite{Spa15,Spa16} showed that the Heisenberg limit can be achieved in a Mach-Zehnder interferometer with squeezed coherent light in both inputs if the squeezing power is roughly $1/3$ of the total power. NOON states \cite{Hol93,Ou96}, have been also shown to reach this limit, however, as pointed out in reference \cite{Cam03}, a parity detection scheme is required.

The previously mentioned papers \cite{Pez08,Lan13,Lan14,Spa15,Spa16,Hol93,Ou96} showed mainly theoretical bounds in the phase sensitivity measurement. Practical detection schemes and their performance \cite{Gar17} as well as the effect of losses \cite{Gar17,Ono10} have to be taken into account when a real-life experiment is described.

This is exactly the approach we shall take in this paper while discussing the MZI phase sensitivity with coherent as well as coherent and squeezed vacuum input light. We focus more on what is experimentally achievable insisting less on optimal but idealistic situations. The detection schemes are described and evaluated while the (high) power of the probe laser beam is not disregarded.

Magnetic field based vacuum birefringence experiments ($B\sim1-10$ T \cite{Zav12,Del14,Del16,Cad14,Mei10}) scrutinize vacuum refraction coefficients in the order of $\Delta{n}\sim10^{-23}$. This extremely small value is self-explanatory why decades of experimental effort failed to give a result. With a $1$-$10$ PW laser we need to detect a refraction index $\Delta{n}\sim10^{-7}-10^{-9}$. Assuming the pump laser to be focused on a few $\mu$m scale, the phase shift induced into a Mach-Zehnder interferometer is in the order of $\Delta\varphi\sim10^{-7}-10^{-9}$ radians. A phase shift measurement at this order of magnitude is challenging, especially if we consider that this phase shift lasts from tens of femto-seconds to tens of pico-seconds (depending on the laser facility), however, it cannot be deemed as infeasible.

Contrary to a common practice in quantum field theories (where $c=1=\hbar$), S.I. units will be employed throughout this paper. The reason for this choice is twofold. First, it avoids confusion among the many (S.I., Gaussian, Heaviside-Lorentz) systems of units found in various works. Second, it gives the final results/formulas to the experimentalists who are not particularly keen on keeping track of various factors that were absorbed in the redefinition of units.


This paper is structured as follows. In Section \ref{sec:VBir_disturbed_vacuum} we discuss the QED theoretical aspects of vacuum birefringence and the predictions for the vacuum refraction index ($n_{QED}$). The proposed experimental setup is detailed in Section \ref{sec:exp_setup}. Maximum theoretical phase sensitivities for four petawatt class laser facilities are computed and discussed in Section \ref{sec:theoretical_feasibility_estimation}. A more realistic approach is taken in Section \ref{sec:signal_detection}, where emphasis is put on the achievable phase sensitivity given the experimental and technical constraints (detection scheme, maximum squeezing, available bandwidth, losses). Section \ref{sec:conclusions} concludes the paper.

\begin{figure}
\centering
\includegraphics[scale=0.65]{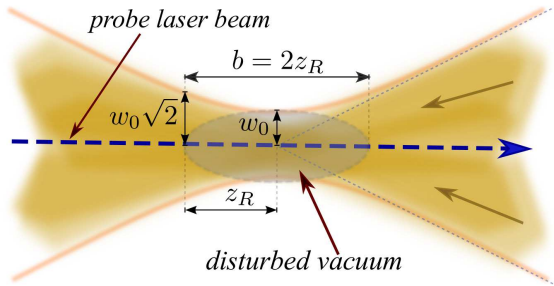}
\caption{\label{fig:VBir_exp_Laser_waist} The pump laser Gaussian beam (propagating from the right) is focused into a region with a waist $w_0$. We have the Rayleigh distance $z_R=\pi{w_0^2}/\lambda_L$ and over the distance $b=2z_R$ we assume the pump laser's electric field reaching its maximum value ${E_L}$. The probe beam (blue dashed line) is to a good approximation counter-propagating in respect with the pump beam.}
\end{figure}

\section{Vacuum birefringence in a disturbed vacuum}
\label{sec:VBir_disturbed_vacuum}
We consider a geometry where the intense linearly polarized petawatt beam (maximum electric (magnetic) field $E_L$ ($B_L=E_L/c$), both perpendicular between them and perpendicular to the direction of propagation) collides almost head-on with the probe beam (see Fig.~\ref{fig:VBir_exp_Laser_waist}). We have the predicted values \cite{Rik00,Rik03,Lui04,Lui04b,Lui05} (see Appendix \ref{sec:app:Heisenberg_Euler})
\begin{equation}
\label{eq:n_para_perp_MO_effect}
\left\{
	\begin{array}{c}
	n_{QED,\parallel}\\
	n_{QED,\perp}
	\end{array}
\right\}
=1+2\xi\times
\left\{
	\begin{array}{c}
	4\\
	7
	\end{array}
\right\}
\times
 E_L^2
\end{equation}
where $n_\parallel$ ($n_\perp$) is the vacuum refraction index when the probe beam's polarization is parallel (perpendicular) to the pump beam's polarization and we have the constant
\begin{equation}
\label{eq:xi_constant_from_QED}
\xi=\frac{\alpha}{45E_S^2}=\frac{\hbar e^4}{180\pi\epsilon_0m_e^4c^7}\approx9.2039\times10^{-41}\quad\frac{\text{m}^2}{\text{V}^2}
\end{equation}
where $\alpha=e^2/4\pi\varepsilon_0\hbar{c}\approx1/137$ is the fine-structure constant and $E_S=m_e^2c^3/e\hbar\approx1.3\times10^{18}$ V/m is the Schwinger-Sauter critical electric field \cite{Sau31,Sch51}. $\hbar$ denotes the reduced Planck constant, $e$ ($m_e$) is the charge (mass) of the electron, $\varepsilon_0$ is the vacuum electric permeability and $c$ is the speed of light in vacuum.

\begin{figure}
\centering
\includegraphics[scale=0.5]{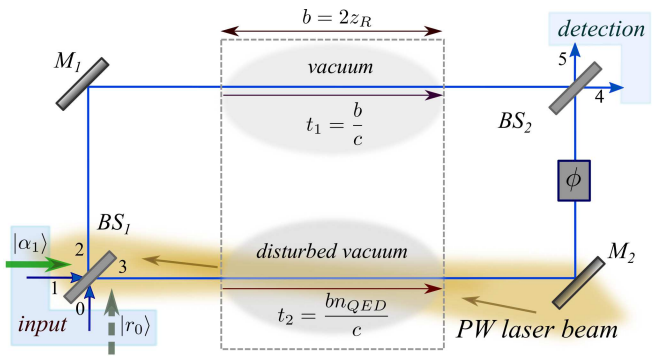}
\caption{\label{fig:VBir_exp_setup} Experimental setup of the proposed all-optical vacuum birefringence experiment. The petawatt-class (pump) laser is counter-propagating and intersecting the probe beam in the lower arm of the interferometer. The QED-induced phase shift $\Delta{\varphi_{QED}}$ needs to be detected at the outputs $4-5$ of the interferometer. The phase shift $\phi$ is chosen by the experimenter in order to reach the MZI's maximum sensitivity.}
\end{figure}

Consider the probe beam (of wavelength $\lambda_p$ and frequency $\omega_p=2\pi{c}/\lambda_p$) counterpropagating over a length $b=2z_R$ (see Fig.~\ref{fig:VBir_exp_Laser_waist}) with a pump laser that disturbed the vacuum . The parameter $b$ is called depth of focus while $z_R=\pi{w_0^2}/\lambda_L$ is called the Rayleigh distance. Here $\lambda_L$ is the pump laser wavelength and $w_0$ the Gaussian beam's waist. The phase shift of this beam in respect with the same beam propagated in an unperturbed vacuum (see Fig.~\ref{fig:VBir_exp_setup}) is
$\Delta\varphi_{QED,\parallel/\perp}={\omega_pb}/{c}\left(n_{QED,\parallel/\perp}-1\right)$
and plugging in the the values from Eq.~\eqref{eq:n_para_perp_MO_effect} yields
\begin{equation}
\label{eq:delta_phi_para_perp}
\left\{
\begin{array}{c}
\Delta\varphi_{QED,\parallel}\\
\Delta\varphi_{QED,\perp}
\end{array}
\right\}
=\frac{8\pi^2w_0^2\xi}{\lambda_p\lambda_L}\times
\left\{
\begin{array}{c}
4\\
7
\end{array}
\right\}
\times E_L^2
\end{equation}
The probe laser beam to have $\lambda_p=532$ nm in our numerical calculations. The pump laser beam with a wavelength $\lambda_L=820$ nm and a waist $w_0\approx3\mu$m. Throughout the depth of focus we assume a constant electrical field (equal to its maximum specified value, $E_L$). For ELI-NP \cite{ELI-NP} and ELI-BL \cite{ELI-BL} this amounts to {${E_L\sim10^{15}}$} V/m yielding the QED-predicted phase shifts
\begin{equation}
\label{eq:delta_phi_para_perp_VALUES}
\left\{
\begin{array}{c}
\Delta\varphi_{QED,\parallel}\\
\Delta\varphi_{QED,\perp}
\end{array}
\right\}
\approx
\left\{
\begin{array}{c}
6\\
10
\end{array}
\right\}
\times 10^{-7}
\end{equation}
For the Vulcan \cite{VULCAN} and XFEL \cite{Miy06} facilities a factor of $10^{-2}$ has to be multiplied to the values from Eq.~\eqref{eq:delta_phi_para_perp_VALUES}.
These predicted phase shifts ($\Delta\varphi\sim10^{-7}$ and, respectively $\Delta\varphi\sim10^{-9}$) are indeed small, however they are not unrealistic to be measured, as we shall show in the following.

\section{Experimental setup}
\label{sec:exp_setup}
Our proposed experimental setup is based on a Mach-Zehnder interferometer \cite{GerryKnight,MandelWolf} driven by a combination of coherent and/or squeezed light (see Fig.~\ref{fig:VBir_exp_setup}.), as we shall shortly discuss. The probe beam interferes in a Mach-Zehnder device formed by the beam splitters $BS_1$ and $BS_2$ (assumed balanced, i. e. transmission / reflection coefficients $T=1/\sqrt{2}$ / $R=i/\sqrt{2}$) and the mirrors $M_1$ and $M_2$. The experimentally adjustable phase shift $\phi$ is used to bring the interferometer to its maximum sensitivity. The counter-propagating petawatt-class laser intersects the lower arm of the interferometer and generates a disturbed vacuum over a distance $b=2z_R$ (i. e. in its focus region). From the probe laser's point of view, this is just a region with a birefringent medium that induces a phase shift. By rotating the polarization of the probe beam (parallel and, respectively, perpendicular to the pump beam) we can measure the QED induced phase shifts. We shall denote in the remainder of the paper the total phase shift by $\varphi$, where $\varphi=\phi+\Delta\varphi_{QED}$ (where $\Delta\varphi_{QED}$ stands for either $\Delta\varphi_{QED,\parallel}$ or $\Delta\varphi_{QED,\perp}$, in function of the probe beam's polarization).

Obviously, the QED induced phase shift has to be larger than the minimum detectable phase sensitivity for our interferometer. It is well known that the phase sensitivity of a Mach-Zehnder is not constant, \cite{Dem15} (unless special techniques are used e. g. \cite{Pez07}). Depending on the input light and detection scenario, one usually has a phase shift $\phi$ that minimizes the phase sensitivity. This topic shall be discussed at large in Sections \ref{sec:theoretical_feasibility_estimation} and \ref{sec:signal_detection}.

In order to remain realistic in our estimations, the petawatt-class (``pump'') laser is assumed to have two types of pulses: 
\begin{enumerate}
	\item[$\bullet$] in the order of tens to hundreds of femto-seconds ($\tau_L\sim10^{-14}$ s for ELI-NP \cite{ELI-NP} and $\tau_L\sim10^{-13}$ s for ELI-BL \cite{ELI-BL})
	\item[$\bullet$] in the order of a fraction to tens of picoseconds ($\tau_L\sim10^{-12}$ s for Vulcan \cite{VULCAN} and $\tau_L\sim10^{-12}-10^{-11}$ s for LFEX \cite{Miy06})
\end{enumerate}
Therefore, the output signal is expected to have a bandwidth $B\sim1/\tau_L$. 

\section{Theoretical phase sensitivity limits}
\label{sec:theoretical_feasibility_estimation}
The challenge in this experiment is the precise measurement of a QED-induced optical phase shift $\Delta{\varphi_{QED}}$ in the MZI. We consider in the following two cases:
\begin{enumerate}
	\item[$\bullet$] classical (coherent) input light
	\item[$\bullet$] non-classical (coherent $\otimes$ squeezed vacuum) input light
\end{enumerate}
We focus in this section on theoretically achievable limits. Experimental issues like the actual detection schemes, signal bandwidth and losses are relegated to Section \ref{sec:signal_detection}.

\subsection{Theoretical phase sensitivity limits using classical light}
\label{subsec:feasibility_estimation_classical}

\subsubsection{CW probe laser}
We first consider an ideal single-mode CW laser applied at input $1$ while input $0$ is kept in the vacuum state. We have the input state vector
\begin{equation}
\label{eq:psi_in_coh_light}
\vert\psi_{in}\rangle=\hat{D}_1\left(\alpha\right)\vert0\rangle
\end{equation}
where the displacement operator \cite{GerryKnight,MandelWolf} acting on input $1$ is defined as $\hat{D}_1\left(\alpha\right)=e^{\alpha\hat{a}_1-\alpha^*\hat{a}_1^\dagger}$. The complex number $\alpha=\vert\alpha\vert{e^{i\theta_\alpha}}$ (not to be confused with the fine structure constant from Section \ref{sec:VBir_disturbed_vacuum} and Appendix \ref{sec:app:Heisenberg_Euler}) denotes the amplitude of the coherent state, $\vert0\rangle$ represents the vacuum state and $\hat{a}_k$ ($\hat{a}_k^\dagger$) denotes the annihilation (creation) operator in mode $k$. For the coherent state under consideration we have the average number of photons $\langle{N}\rangle=\langle\psi_{in}\vert\hat{a}_1^\dagger\hat{a}_1\vert\psi_{in}\rangle=\vert\alpha\vert^2$.

The CW probe laser beam has an energy per light quantum $\epsilon_p=\hbar\omega_p=hc/\lambda_p\approx3.7\times10^{-19}$ J. The average number of photons in the interferometer over the relevant timescale (i.e. the pump laser pulse duration $\tau_L$) is given by
\begin{equation}
\label{eq:N_average}
\langle{N}\rangle=\frac{P\tau_L}{\hbar\omega_p}=\frac{P\tau_L}{\epsilon_p}
\end{equation}

It is common knowledge that the minimum phase measurement error of a standard MZI using coherent light is lower bounded by the standard quantum limit \cite{Dem15}
\begin{equation}
\label{eq:Delta_phi_SQL}
\Delta\varphi_{SQL}\geq\frac{1}{\sqrt{\langle{N}\rangle}}=\frac{1}{\vert\alpha\vert}
\end{equation}
From Table~\ref{tab:Laser_parameters_and_sensitivities_COH}, it is clear that the maximum realistic powers ($P\sim100-500$ W) for the CW probe beam are insufficient to detect the QED predicted phase shift $\Delta\varphi_{QED}$.  Eq.~\eqref{eq:Delta_phi_SQL} also allows one to compute the minimum power for the probe laser in order to satisfy the QED-predicted phase error ($\Delta\varphi_{QED}$), namely
\begin{equation}
\label{eq:P_average_coherent_given_Delta_phi}
P\approx\frac{\epsilon_l}{\tau_L\left(\Delta\varphi_{QED}\right)^2}
\end{equation}
and plugging in the values from Eq.~\eqref{eq:delta_phi_para_perp_VALUES} implies immediately $P\sim10^{10}$ W, a totally unrealistic power for a CW laser.

\subsubsection{Pulsed probe laser}
\label{subsec:Pulsed_laser_theo}
However, such (peak) powers are commonplace in table-top pulsed lasers. Powers in the order of {${P\sim10^{12}}$~W} are readily available \cite{Bar94,Bar96}, pushing the theoretical phase sensitivity orders of magnitude below the required level. A few terawatt femto-/pico-second class laser (i. e. probe pulse duration $\tau_p\sim10^{-14}-10^{-10}$ s) implies an average number of photons over the relevant pump laser timescale (from $\tau_L\sim10^{-14}$ to $\tau_L\sim10^{-12}$ s) of $\langle{N}\rangle\sim10^{16}-10^{18}$ photons implying a sensitivity $\Delta\varphi\sim10^{-8}-10^{-9}$, enough to detect de QED signal. These results are summarized in the last two columns of Table~\ref{tab:Laser_parameters_and_sensitivities_COH}.

The main challenge in this scenario is the synchronisation with the pump laser (accuracy needed $\sim10^{-14}-10^{-11}$ s, depending on the pump laser pulse) and the phase stability of the (pulsed) probe laser.

\subsection{Theoretical phase sensitivity limits using non-classical light}
\label{subsec:feasibility_estimation_quantum}
One can add squeezed vacuum into the unused port $0$, therefore the input state vector can be written as
\begin{equation}
\label{eq:psi_non_coh_squeezed_vac}
\vert\psi_{in}\rangle=\vert{r_0}\alpha_1\rangle=\hat{S}_0\left(r\right)\hat{D}_1\left(\alpha\right)\vert0\rangle
\end{equation}
with the squeeze operator \cite{GerryKnight} $\hat{S}_0\left(r\right)=e^{r/2\left(\hat{a}_0^2-\left(\hat{a}_0^\dagger\right)^2\right)}$ where $r\in\mathbb{R}^+$ is the squeezing factor. The best achievable phase sensitivity measurement with this type of input has been shown to be \cite{Pez08,Lan13,Lan14}
\begin{equation}
\label{eq:Delta_phi_coherent_squeezed}
\Delta\varphi_{CSV}\geq\frac{1}{\sqrt{\vert\alpha\vert^2e^{2r}+\sinh^2r}}
\end{equation}
The phase sensitivity $\Delta\varphi_{CSV}$ from Eq.~ \eqref{eq:Delta_phi_coherent_squeezed} reaches the Heisenberg limit 
\cite{Dem15,Pez08,Lan13,Lan14,Spa15,Spa16,Hol93,Ou96},
\begin{equation}
\label{eq:Delta_phi_Heisenberg_limit}
\Delta\varphi_{HL}\approx\frac{1}
{\langle{N}\rangle}
\end{equation}
only if  we impose \cite{Pez08}
\begin{equation}
\label{eq:CSV_condition_for_HL}
\vert\alpha\vert^2\approx\sinh^2r\approx\frac{\langle{N}\rangle}{2}
\end{equation}
In Table~ \ref{tab:Laser_parameters_and_sensitivities_SQZ}, the QED predicted phase shift  and the Heisenberg-limited sensitivity for our MZI are given for four petawatt-class lasers (ELI-NP \cite{Hom16}, ELI-BL \cite{ELI-BL}, Vulcan \cite{VULCAN} and LFEX \cite{Miy06}).

Using a CW coherent probe beam in port $1$ with only ${P=10}$ W plus squeezing in port  $0$ (see Fig.~\ref{fig:VBir_exp_Laser_waist}) so that the phase sensitivity becomes Heisenberg-limited \eqref{eq:Delta_phi_Heisenberg_limit}, two facilities (ELI-BL and LFEX) are theoretically feasible for this experiment. If we push the power of our probe laser to $P\sim100$ W (similar to the one used by LIGO) and somehow manage to maintain the squeezing factor constraint \eqref{eq:CSV_condition_for_HL}, a simple calculation shows that all considered facilities are theoretically within the sensitivity range to detect the QED predicted effect. We shall critically discuss this scenario in Section \ref{subsec:detection_non_classical_light}.

If one cannot satisfy the Heisenberg scaling condition~\eqref{eq:CSV_condition_for_HL} and we assume $\vert\alpha\vert^2\gg\sinh^2r$ (i. e. the available laser power is much higher than the squeezing), then we can approximate the minimum phase uncertainty with
\begin{equation}
\label{eq:Delta_phi_coherent_squeezed_approx}
\Delta\varphi_{CSV}\approx\frac{e^{-r}}{\vert\alpha\vert}=\frac{e^{-r}}{\sqrt{\langle{N}\rangle}}
\end{equation}
The squeezing still brings an $e^{-r}$ gain in the phase sensitivity, i.e. {${\Delta\varphi_{CSV}\approx e^{-r}\Delta\varphi_{SQL}}$}. In today's technology this would mean an order of magnitude \cite{Vah16}. We shall discuss this case at large in Section \ref{subsec:detection_non_classical_light}, too.

Another proposal employs squeezing in both ports. Instead of using the coherent $\otimes$ squeezed input state configuration, Sparaciari, Olivares and Paris considered squeezed coherent states in both inputs \cite{Spa15,Spa16}. Therefore, we consider the input state
\begin{equation}
\vert\psi_{in}\rangle=\vert\gamma,\zeta\rangle_0\otimes\vert\alpha,\xi\rangle_1
\end{equation}
where $\vert{a,b}\rangle_j=\hat{D}_j\left(a\right)\hat{S}_j\left(b\right)\vert0\rangle$, $a\in\{\alpha,\gamma\}$, $b\in\{\zeta,\xi\}$ and $j=0,1$. In \cite{Spa15} it is shown that the maximum Fisher information (and thus the best phase sensitivity) is achieved when $\alpha=\gamma$ and $\xi=\zeta=r$ (all taken real in their paper). In other words, the optimum case is when both input lasers have the same power and both squeezings are identical. If we denote the total squeezing factor $\beta_{tot}=2\sinh^2r/N_{tot}$ where $N_{tot}=2\left(\vert\alpha\vert^2+\sinh^2r\right)$ is the total number of photons inside the interferometer, we have the phase sensitivity given by \cite{Spa15}
\begin{equation}
\label{eq:Delta_squeezed_coh_Sparaciari_Paris}
\Delta\varphi_{SQC}\geq\frac{1}{\sqrt{\frac{8N_{tot}^2\left(2+\sqrt{1+3N_{tot}^{-1}}\right)}{9}+4N_{tot}}}
\approx\frac{1}{\frac{4}{3}N_{tot}}
\end{equation}
where in this case $\beta_{tot}=2/3$ and we assumed a large energy regime ($N_{tot}\gg1$). This result leads indeed to a Heisenberg scaling with an improved proportionality constant (compared to the coherent $\otimes$ squeezed vacuum case).

\section{Realistic phase sensitivity estimation}
\label{sec:signal_detection}
It is self-understood that the interferometer can be operated in vacuum only. Due to the smallness of the expected phase shift, seismic isolation and environmental vibration damping have to be considered, too. The mirrors and beam splitters from Fig.~\ref{fig:VBir_exp_setup} should be suspended by double pendulums \cite{Aso04} or using stabilized platforms with actuators \cite{Num08}.

In order not to average out the QED-induced signal, the photo-detector(s) should be triggered (with a suitable delay) after the moment of interaction of the probe beam with the pump laser pulse. The integration time of the photo-detector should also be matched with the pump pulse duration. This is not a problem for the pico-second pump pulses however it is rather problematic for femto-second one. Depending on the experimental scenario, more complex detection schemes might be considered  \cite{Sch11}.

\subsection{Classical light input}
\label{subsec:detection_classical_light}

\subsubsection{CW probe laser}
The sensitivity of a Mach-Zehnder given by Eq.~\eqref{eq:Delta_phi_SQL} is a lower bound. If we use as an observable the difference in photo-counts $\hat{N}_d=\hat{a}_4^\dagger\hat{a}_4-\hat{a}_5^\dagger\hat{a}_5$ at detectors $D_4$ and $D_5$ (see Fig.~\ref{fig:MZI_2D_detection_schemes}), one actually has (see Appendix \ref{sec:app:sensitivity_calculations}) the phase sensitivity
\begin{equation}
\label{eq:Delta_varphi_coherent_sin_theta}
\Delta\varphi=\frac{1}{\vert\alpha\vert\vert\sin\varphi\vert}
=\frac{1}{\sqrt{\langle{N}\rangle}\vert\sin\varphi\vert}
\end{equation}
where we remind that  $\varphi=\Delta\varphi_{QED}+\phi$. Since ${\Delta\varphi_{QED}\ll1}$, the optimum point is when the experimenter imposes $\phi=\pi/2$. This scenario implies large signals on both detectors being impractical for sensitive photo-detectors PIN diodes. (The highest efficiency for PIN diodes is in the $\mu$W range \cite{Bro10}.)

\begin{figure}
\centering
\includegraphics[scale=0.55]{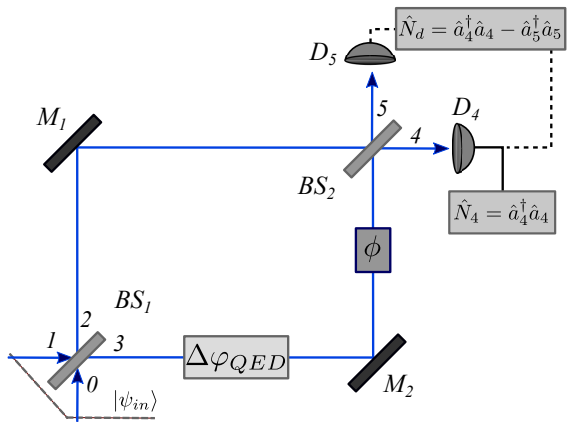}
\caption{\label{fig:MZI_2D_detection_schemes} The actual detection schemes used for our setup. The ``differential detection'' subtracts the photo-currents from the detectors $D_4$ and $D_5$, therefore the corresponding operator is $\hat{N}_d$. The second option is to use a single detector and we chose $D_4$, therefore the relevant operator is $\hat{N}_4$. The experimenter-controlled phase shift $\phi$ brings the Mach-Zehnder to its optimum phase sensitivity so that that the very small QED-induced phase shift $\Delta{\varphi_{QED}}$ can be detected.}
\end{figure}

A single detector setup employing the observable ${\hat{N}_4=\hat{a}_4^\dagger\hat{a}_4}$ (see Fig.~\ref{fig:MZI_2D_detection_schemes}) yields a phase sensitivity (details are given in Appendix \ref{sec:app:sensitivity_calculations})
\begin{equation}
\label{eq:delta_phi_N4_coherent}
\Delta\varphi=\frac{1}{\sqrt{\langle{N}\rangle}\big\vert\sin\left(\frac{\varphi}{2}\right)\big\vert}
\end{equation}
This time the optimum is when $\varphi=\pi$ and this implies a very small signal at the detector $D_4$ (``dark port''). This is a desirable scenario since a high-efficiency PIN diode can be used as a detector \cite{Bro10}.

\subsubsection{Pulsed probe laser}
As discussed in Section \ref{subsec:feasibility_estimation_classical}, a pulsed laser has orders of magnitude higher peak powers compared to its CW counterpart, therefore a noticeable improvement in the phase sensitivity can be expected.
 
Indeed, quantum-limited measurements with a femto-second pulsed laser have been reported \cite{Jia13,Thi17} and their scaling yields
\begin{equation}
\Delta\varphi\sim\frac{1}{2\sqrt{N}\sqrt{\frac{\Delta\omega^2}{\omega_l^2}+1}}
\end{equation}
where $\Delta\omega$ is the spectral width of the pulsed probe laser. Hence, the SQL scaling should be practically achievable for the pulsed probe laser, too.

To the synchronization between the probe laser and the detector trigger discussed in the beginning of this section, one has to add now the synchronization in order to obtain a ``head-on'' collision of the probe and the pump pulses at the focus point of the pump laser. This synchronization requirement is in the order of $\tau_L$. For the femto-second laser facilities \cite{ELI-NP,ELI-BL} this is a serious issue. It is less of a problem for the picosecond pump pulses \cite{VULCAN,Miy06}.

\subsection{Non-classical light input}
\label{subsec:detection_non_classical_light}
In Section \ref{subsec:feasibility_estimation_quantum} we concluded that if we arrive at a Heisenberg scaling even a CW laser with $P=10$ W could provide enough sensitivity to detect the vacuum birefringence signal.
 While theoretically possible, at a closer look satisfying Eq.~\eqref{eq:CSV_condition_for_HL} implies a squeezing factor beyond today's technological possibilities.

The current record for the squeezing factor ($15.3$ dB or $r=1.7$) is reported in \cite{Vah16}. For example, a probe laser power $P=10$ W would imply $\langle{N}\rangle\approx4\times10^6$ photons during the relevant timescale for ELI-BL, therefore according to Eq.~\eqref{eq:CSV_condition_for_HL} a value of $r\approx8$ would be needed to attain the Heisenberg limit. This is an unrealistic value from an experimental point of view.

The remark previously stated holds for the squeezed coherent scenario \cite{Spa15,Spa16}, too: the high squeezing factors (i. e. the $\beta_{tot}$ factor in this case) needed to follow a Heisenberg scaling for powers in the Watt range (and above) are technically difficult today.

We note, nonetheless, that there is \emph{no theoretical limitation} on value of $r$ (or $\beta_{tot}$), the experimental limitations arising mainly from the losses in the generation/detection of this state \cite{Sch17}. The rapid progress of squeezing in the last decade ($7$ dB ($r=0.8$) in 2007, $10$ dB ($r=1.15$) in 2010 \cite{Vah10}, $15$ dB in 2016 \cite{Vah16}) suggests further technological improvements \cite{Sch17}. We can speculate that a $20$ dB ($r=2.3$) squeezing is within reach in the near future and a more optimistic $25-30$ dB ($r=2.8-3.5$) squeezing is not excluded in the next decade.

In the last two columns of Table~\ref{tab:Laser_parameters_and_sensitivities_SQZ} we compute the achievable phase sensitivities for a coherent $\otimes$ squeezed vacuum input light. We take our probe CW laser with $P=200$ W and consider two squeezing factors: ${r=3.5}$ (optimistic, however foreseeable in the future) and ${r=6}$ (rather idealistic). One notes that roughly one order of magnitude is missing in order to meet the required QED phase sensitivity for $r=3.5$ and for the rather too optimistic squeezing factor $r=6$ despite not being Heisenberg-limited, the sensitivity is reached for the ELI lasers.

In the following we consider the realistic detection schemes depicted in Fig.~\ref{fig:MZI_2D_detection_schemes} and evaluate their performance. For a difference detection scheme, the phase sensitivity of a coherent $\otimes$ squeezed vacuum input is given by \cite{Dem15,Gar17} (see also Appendix \ref{sec:app:sensitivity_calculations})
\begin{eqnarray}
\label{eq:Delta_phi_d_coh_sqz_vac_diff}
\Delta\varphi
=\frac{\sqrt{\frac{\vert\alpha\vert^2+\frac{\sinh^22r}{2}}{\tan^2\varphi}
+\sinh^2r+\vert\alpha\vert^2e^{-2r}}}
{\big\vert\vert\alpha\vert^2-\sinh^2r\big\vert}
\end{eqnarray}
and for $\varphi=\pi/2$ we have the optimum phase sensitivity
\begin{eqnarray}
\label{eq:Delta_phi_d_coh_sqz_opt_phi}
\Delta\varphi
=\frac{\sqrt{\sinh^2r+\vert\alpha\vert^2e^{-2r}
}}{\big\vert\vert\alpha\vert^2-\sinh^2r\big\vert}
\end{eqnarray}
In our setup we have $\vert\alpha\vert^2\gg\sinh^2r$ and imposing this condition to Eq.~\eqref{eq:Delta_phi_d_coh_sqz_opt_phi} leads to the result from Eq.~\eqref{eq:Delta_phi_coherent_squeezed_approx}. Thus, we are not Heisenberg limited but we gain an $e^{-r}$ factor compared to the SQL.

However, a total phase shift $\varphi=\pi/2$ inside the MZI implies equal output powers at the detectors $D_4$ and $D_5$ causing additional problems for a high-efficiency detection in the case of a high-power probe beam. We thus focus on the single-detection case. If we consider detection events solely at $D_4$, we arrive at a phase sensitivity (see calculations in Appendix \ref{sec:app:sensitivity_calculations})
\begin{eqnarray}
\label{eq:Delta_phi_sqz_coh_SINGLE_DET_theta_zero}
\Delta\varphi
=\frac{\sqrt{\frac{\sinh^22r\tan^2\left(\frac{\varphi}{2}\right)}{2}
+\sinh^2r
+\frac{\vert\alpha\vert^2}{\tan^2\left(\frac{\varphi}{2}\right)}
+\frac{\vert\alpha\vert^2}{e^{2r}}
}}
{\Big\vert \vert\alpha\vert^2-\sinh^2r\Big\vert}
\end{eqnarray}
An optimum point can be found for the total phase shift inside the MZI
\begin{equation}
\label{eq:phi_optimal_CSV_single_DET}
\varphi_{opt}=\pm2\arctan\left(\sqrt{\frac{\sqrt{2}\vert\alpha\vert}{\sinh2r}}\right)+2k\pi
\end{equation}
with $k\in\mathbb{Z}$ and introducing this value into Eq.~\eqref{eq:Delta_phi_sqz_coh_SINGLE_DET_theta_zero} gives the best achievable sensitivity in the case of a single detector scheme namely
\begin{eqnarray}
\label{eq:Delta_phi_sqz_coh_SINGLE_DET_OPTIMAL}
\Delta\varphi
=\frac{\sqrt{\sinh^2r+\sqrt{2}\vert\alpha\vert\sinh2r+\vert\alpha\vert^2e^{-2r}}}
{\big\vert\vert\alpha\vert^2-\sinh^2r\big\vert}
\end{eqnarray}
result also found in reference \cite{Gar17}. For a strong coherent source and a realistic squeezing factor we have $\vert\alpha\vert^2\gg\vert\alpha\vert$ and $\vert\alpha\vert\gg\sinh{r}$, therefore the phase sensitivity from Eq.~\eqref{eq:Delta_phi_sqz_coh_SINGLE_DET_OPTIMAL}, similar to the difference detection scheme, migrates to Eq.~\eqref{eq:Delta_phi_coherent_squeezed_approx}. It is noteworthy to point out that we have again a ``dark port'' scenario ($D_4$ gets almost no light) and this is the optimum situation for sensitive photo-detectors. Thus, for high input powers, the single detector scheme is preferable to the difference detection setup.

\subsection{Detection bandwidth}
\label{subsec:detection_bandwidth}
The fact that the useful signal at the outputs $4$ and $5$ lasts only for a short time (depending on the laser facility, from dozens of femtoseconds to tens of picoseconds) implies a detection scheme having the necessary bandwidth to cope with this signals.

For coherent light, the generation \cite{Bar94} and detection \cite{Sch11} of femto-second pulses do not pose a problem. This assertion is no longer true for squeezed states. (For example adding in the detection scheme beam splitters with one input port ``empty'' will introduce extra vacuum fluctuations, thus reducing or even cancelling the advantage brought by squeezed light.)

The typical scheme used (see, e. g. Fig.~20 in reference \cite{Sch17}) to generate squeezed vacuum is based on a SHG (second harmonic generation) followed by a sub-threshold pumped OPA (optical parametric amplifier) \cite{Bre98,Aba11,Vah10,Vah16}. Among the reasons for placing the non-linear crystal in an OPA cavity are an enhanced down-conversion efficiency and confining the transversal mode (usually $\text{TEM}_{00}$). The output spectrum of the squeezed ($\Delta^2\hat{X}_+$) and anti-squeezed ($\Delta^2\hat{X}_-$) quadrature variances are given by 
\begin{equation}
\label{eq:Delta2_X_squeeze_antisqueeze}
\Delta^2\hat{X}_{\pm}\left(\Omega\right)=1\pm\eta_{d}\frac{4\sqrt{p}}
{\left(1\mp\sqrt{p}\right)^2+\left(\Omega/\Gamma\right)^2}
\end{equation}
where $p=P/P_{th}$, $P$ is the second harmonic pump power, $P_{th}$ is the OPA threshold power, $\Omega$ is the sideband frequency and $\eta_{d}$ denotes all optical losses. The cavity linewidth (also called decay rate) is $\Gamma=c\left(T+L\right)/l$ where $l$ is the cavity length, $T$ denotes the cavity mirror's power transmission coefficient and $L$ is the round-trip loss.

For the detection of the vacuum birefringence signal created by a pump laser pulse of duration $\tau_L$ we need a bandwidth in the order of $B\sim1/\tau_L$. This would imply for example for the LFEX laser a minimum of $B\sim50$ GHz and for the ELI-BL laser $B\sim6.7$ THz.

The ``natural bandwidth'' of the OPA generated squeezed states is in the tens to hundreds of megahertz \cite{Bre98,Vah10,Vah16}. It is here that the record in squeezing has been achieved. However, ultra-broadband squeezed light ($B\sim13.4$ THz in the telecommunication band $\lambda=1535$) has been reported \cite{Wak14}. Therefore, bandwidth does not seem to pose a particular problem to squeezed states, the technological problem to answer is the amount of squeezing achievable in the ultra-broadband case.

\subsection{The impact of losses on the phase sensitivity}
\label{subsec:detection_losses}
All results from discussed up to this point assume no losses. In some scenarios this might not be realistic. 

Following \cite{Ono10}, we can define $\sigma$ the loss ratio i. e. the probability that any photon is lost between  generation and detection. Then, for a coherent state we have to make the replacement $\alpha'\to\sqrt{1-\sigma}\alpha$ resulting in a bound
\begin{equation}
\label{eq:Delta_varphi_losses_coherent}
\Delta\varphi_{SQL}^l=\frac{1}{\sqrt{1-\sigma}\vert\alpha\vert}
=\frac{\Delta\varphi_{SQL}}{\sqrt{1-\sigma}}
\end{equation}
Especially for low losses (i. e. $\sigma\ll1$) we have a small effect for the phase sensitivity. Therefore, in an experimental setup employing only classical light, losses are not of paramount importance.

For the coherent $\otimes$ squeezed vacuum case we can approximate \cite{Ono10} the phase sensitivity
\begin{equation}
\Delta\varphi_{CSV}^l
\approx\frac{\sqrt{\sigma+(1-\sigma)e^{-2r}}}{\sqrt{\left(1-\sigma\right)\vert\alpha\vert^2+\sigma\left(1-\sigma\right)\sinh^2r}}
\end{equation}
The effect of losses is more vicious for high squeezing factors. Indeed, for  $\sigma\gg e^{-2r}$, we arrive at the SQL scaling given by Eq.~\eqref{eq:Delta_varphi_losses_coherent} for $\vert\alpha\vert^2\gg\sinh^2r$ and the whole interest of squeezing is lost.

For the squeezed coherent case \cite{Spa15}, too, it has been shown in \cite{Spa16} that including losses degrades the performance from a Heisenberg scaling to a shot-noise scaling. A lengthier discussion about the effect of losses can be found the literature \cite{Gar17,Ono10,Dem12}.

\subsection{Repeated measurements}
\label{subsec:baey}
All results discussed up to this point assumed a single measurement. For $N_{exp}$ independent measurements having the same probability distribution of the measured observable, the Fisher information is additive \cite{Dem15} resulting in $F\left(\varphi\right)\to N_{exp}F\left(\varphi\right)$. The Cram\'er-Rao lower bound is therefore given by
\begin{equation}
\label{eq:Delta_phi_multi_Delta_phi_single}
\Delta\varphi_{\textrm{multi}}=\frac{1}{\sqrt{N_{exp}}}\Delta\varphi_{\textrm{single}}
\end{equation}
The scaling with $1/\sqrt{N_{exp}}$ has been shown both theoretically and experimentally for coherent light \cite{Pez07} and for coherent and squeezed vacuum \cite{Pez08}.

While theoretically the limit $N_{exp}\to\infty$ can be taken, in realistic experiments many constraints limit the acquisition time. In a high-power laser facility, one of the most stringent constraints is the number of PW laser shots available for a given experiment. Realistically we can estimate this number between dozens and thousands, depending on the laser facility.

With an order of magnitude sensitivity gain from repeated measurements and data processing, the figures from Tables \ref{tab:Laser_parameters_and_sensitivities_COH} and \ref{tab:Laser_parameters_and_sensitivities_SQZ} become more optimistic with feasible scenarios including e. g. the CW $P=200$ W probe beam and a squeezing factor of $r=3.5$.

\subsection{Boosting the phase sensitivity with SU(1,1) interferometers}
On a more speculative note, one can also consider the active interferometers setup \cite{Spa15,Spa16} also called SU(1,1) interferometer where the beam splitters ($BS_1$ and/or $BS_2$) are replaced by non-linear crystals acting as OPAs. These types of interferometers have a long history, being first discussed in the 1980s (Wodkiewicz and Eberly \cite{Wod85}, Yurke et al. \cite{Yur86}). Theoretical phase sensitivity studies \cite{Li14} as well as experimental demonstrations \cite{Jin11,And17} confirmed the better performance of this setup versus the passive Mach-Zehnder, especially when losses are involved. The main drawback of the active SU(1,1) interferometer is its very low operational power, therefore for our purpose it seems not feasible for the moment. Further improvements in its operational power could make it an interesting candidate in the future.

\section{Conclusions}
\label{sec:conclusions}
In this paper we introduced and discussed an all-optical experimental setup for the detection of the QED-predicted vacuum birefringence. The proposal is based on a Mach-Zehnder interferometer operated between the standard quantum limit and the Heisenberg limit through the use of classical and non-classical states of light. The vacuum birefringence signal is created by a petawatt class laser and its signature is a delay (phase shift) detectable at the output of the interferometer.

Multi-petawatt class lasers can induce a significant delay on a counter-propagating probe laser beam, therefore its detection becomes -- despite the technological difficulties -- foreseeable. Various scenarios have been shown to be feasible, including terawatt pulsed lasers for the probe beam and the use of squeezed states of light.

We conclude by remarking that this experiment does not ``consume'' at all power from the pump laser. It can be placed upstream, immediately after the compressor of the high power laser system. The petawatt laser pulse, after disturbing the quantum vacuum inside the Mach-Zehnder interferometer, can be used for other downstream experiments.

\section{Acknowledgement}

The author acknowledges financial support from the Extreme Light Infrastructure Nuclear Physics (ELI-NP) Phase II, a project co-financed by the Romanian Government and the European Union through the European Regional Development Fund.

\appendix
\section{Derivation of the vacuum refraction index}
\label{sec:app:Heisenberg_Euler}
The effective QED Lagrangian including non-linearities at all orders (i. e. the Heisenberg-Euler Lagrangian, see e. g. Eq.~(14) in \cite{Bat12}) can be expanded in its lowest order non-linear contributions yielding the so-called Euler-Kockel Lagrangian \cite{Eul35}
\begin{eqnarray}
\label{eq:EK_Lagrangian}
\mathcal{L}_{EK}=\frac{\mathcal{F}}{2}+\kappa\left(\mathcal{F}^2+7\mathcal{G}^2\right)
\end{eqnarray}
where introduced the relativistic invariants ${\mathcal{F}=\varepsilon_0/2\left(E^2-c^2B^2\right)}$, ${\mathcal{G}=\sqrt{\varepsilon_0/\mu_0}\left(\boldsymbol{E}\cdot\boldsymbol{B}\right)}$ and $X^2=\boldsymbol{X}\cdot\boldsymbol{X}$ where $\boldsymbol{X}\in\{\boldsymbol{E},\boldsymbol{B}\}$. Also, for compactness we made the notation $\kappa=2\alpha^2\hbar^3/45m_e^4c^5$. One can now write the constitutive relations from the above Lagrangian i. e. $\boldsymbol{D}_{EK}=\partial\mathcal{L}_{EK}/\partial\boldsymbol{E}$ and $\boldsymbol{H}_{EK}=-\partial\mathcal{L}_{EK}/\partial\boldsymbol{B}$ yielding
\begin{equation}
\label{eq:app:constitutive_eqs_Euler_Kockel}
\left\{
\begin{array}{l}
\boldsymbol{D}_{EK}=\epsilon_0\left(1+4\kappa\mathcal{F}\right)\boldsymbol{E}+14\epsilon_0c\kappa\mathcal{G}\boldsymbol{B}
\\
\boldsymbol{H}_{EK}=\frac{1}{\mu_0}\left(1+4\kappa\mathcal{F}\right)\boldsymbol{B}-14\epsilon_0c\mathcal{G}\boldsymbol{E}
\end{array}
\right.
\end{equation}
The total electric field is the (vectorial) sum of the probe ($\boldsymbol{E}_p$) and pump electric fields $\boldsymbol{E}=\boldsymbol{E}_p+\boldsymbol{E}_L$ and the same goes for the magnetic field, $\boldsymbol{B}=\boldsymbol{B}_p+\boldsymbol{B}_L$. We use the relation $E_p^2-c^2B_p^2=0$ and retain from Eq.~\eqref{eq:app:constitutive_eqs_Euler_Kockel} only the components at the probe frequency (i. e. linear in $\boldsymbol{E}_p$ and $\boldsymbol{B}_p$). Next, following reference \cite{Rik00}, we use Maxwell's equations in the classical limit and assume that the monochromatic probe field ``sees'' a refraction index $n_{QED}$ (to be determined). Combining the previous findings allows one to write an equation whose eigenvalues yield the refraction indices \cite{Rik00}. In the case of two, counter-propagating light beams, one finds the parallel/perpendicular indices of refraction \cite{Rik00,Lui05} seen by the probe beam
\begin{equation}
\left\{
\begin{array}{l}
n_{QED,\parallel}=1+2\xi\left(E_L^2+2cE_LB_L+c^2B_L^2\right)\\
n_{QED,\perp}=1+\frac{7}{2}\xi\left(E_L^2+2cE_LB_L+c^2B_L^2\right)
\end{array}
\right.
\end{equation}
and since we have for the pump laser beam $E_L=cB_L$ we arrive at the expression from Eq.~\eqref{eq:n_para_perp_MO_effect}.

We mention that other, more elaborate, methods to compute the vacuum refraction indices exist, mainly based on the polarization tensor \cite{Bec91,Dit85}.

\section{Short theoretical discussion of the MZI phase sensitivity}
\label{sec:app:sensitivity_calculations}
Let a Hermitian operator $\hat{O}$ depend on a parameter we wish to measure, say, $\varphi$. We can measure the average of this operator as
$\langle\hat{O}\left(\varphi\right)\rangle=\langle\psi\vert\hat{O}\left(\varphi\right)\vert\psi\rangle$. A small variation $\delta\varphi$ of the parameter $\varphi$ causes a change
\begin{equation}
\langle\hat{O}\left(\varphi+\delta\varphi\right)\rangle\approx\langle\hat{O}\left(\varphi\right)\rangle+\frac{\partial\langle\hat{O}\left(\varphi\right)\rangle}{\partial\varphi}\delta\varphi
\end{equation}
thus we have the approximative result
${\langle\hat{O}\left(\varphi+\delta\varphi\right)\rangle-\langle\hat{O}\left(\varphi\right)\rangle
\approx\frac{\partial\langle\hat{O}\left(\varphi\right)\rangle}{\partial\varphi}\delta\varphi}$. In order to detect this difference of averages, we have to require that
\begin{equation}
\label{eq:delta_O_operators_greater_variance_O}
\vert\langle\hat{O}\left(\varphi+\delta\varphi\right)\rangle-\langle\hat{O}\left(\varphi\right)\rangle\vert\geq\Delta\hat{O}\left(\varphi\right)
\end{equation}
where the standard deviation $\Delta\hat{O}$ of the operator $\hat{O}$ is defined as the square root of its variance, ${\Delta^2\hat{O}=\langle\hat{O}^2\rangle-\langle\hat{O}\rangle^2}$. The value of $\delta\varphi$ that saturates inequality \eqref{eq:delta_O_operators_greater_variance_O} is called sensitivity (denoted $\Delta\varphi$ throughout this paper) and is given by
\begin{equation}
\label{eq:Delta_varphi_DEFINITION}
\Delta\varphi=\frac{\Delta\hat{O}}{\big\vert\frac{\partial}{\partial\varphi}\langle\hat{O}\rangle\big\vert}
\end{equation}
In the difference detection scheme (see Fig.~\ref{fig:MZI_2D_detection_schemes}), we have the observable $\hat{N}_d=\hat{a}_4^\dagger\hat{a}_4-\hat{a}_5^\dagger\hat{a}_5$. Using standard field operator transformations in a balanced MZI \cite{GerryKnight}
\begin{equation}
\label{eq:field_op_transf_MZI}
\left\{
\begin{array}{l}
\hat{a}_4^\dagger=-\sin\left(\frac{\varphi}{2}\right)\hat{a}_0^\dagger+\cos\left(\frac{\varphi}{2}\right)\hat{a}_1^\dagger\\
\hat{a}_5^\dagger=\mbox{ }\cos\left(\frac{\varphi}{2}\right)\hat{a}_0^\dagger+\sin\left(\frac{\varphi}{2}\right)\hat{a}_1^\dagger
\end{array}
\right.
\end{equation}
(some global phase factors have been ignored) we get
\begin{equation}
\hat{N}_d=\cos\varphi\left(\hat{a}_1^\dagger\hat{a}_1-\hat{a}_0^\dagger\hat{a}_0\right)-\sin\varphi\left(\hat{a}_1^\dagger\hat{a}_0
+\hat{a}_1\hat{a}_0^\dagger\right)
\end{equation}
In the case of the input state vector from Eq.~\eqref{eq:psi_in_coh_light} we immediately have $\vert\partial\langle\hat{N}_d\rangle/\partial\varphi\vert=\vert\sin\varphi\vert\cdot\vert\alpha\vert^2$ and the variance yields $\Delta^2\hat{N}_d=\vert\alpha\vert^2$. Applying these results to Eq.~\eqref{eq:Delta_varphi_DEFINITION} gives $\Delta\varphi$ from Eq.~\eqref{eq:Delta_varphi_coherent_sin_theta}. With a single detector scheme we have the observable $\hat{N}_4=\hat{a}_4^\dagger\hat{a}_4$ which, after performing the MZI field operator transformations given by Eq.~\eqref{eq:field_op_transf_MZI} yields
\begin{eqnarray}
\hat{N}_4=\sin^2\left(\frac{\varphi}{2}\right)\hat{a}_0^\dagger\hat{a}_0+\cos^2\left(\frac{\varphi}{2}\right)\hat{a}_1^\dagger\hat{a}_1
\nonumber\\
-\frac{\sin\varphi}{2}\left(\hat{a}_1^\dagger\hat{a}_0
+\hat{a}_1\hat{a}_0^\dagger\right)
\end{eqnarray}
thus ${\vert{\partial\langle\hat{N}_4\rangle}/{\partial\varphi}\vert
=\vert\sin\varphi\vert\vert\alpha\vert^2/2}$ and the variance is ${\Delta^2\hat{N}_4=\cos^2\varphi\vert\alpha\vert^2}$.  Plugging these results into Eq.~\eqref{eq:Delta_varphi_DEFINITION} yields $\Delta\varphi$ from Eq.~\eqref{eq:delta_phi_N4_coherent}.

In the case of a coherent $\otimes$ squeezed vacuum input state given by Eq.~\eqref{eq:psi_non_coh_squeezed_vac}, for a differential detection scheme we have 
\begin{equation}
\label{eq:del_Nd_average_over_del_phi_coh_sq}
\bigg\vert\frac{\partial\langle\hat{N}_d\rangle}{\partial\varphi}\bigg\vert
=\vert\sin\varphi\vert\cdot\big\vert\vert\alpha\vert^2-\sinh^2r\big\vert
\end{equation}
and a variance
\begin{eqnarray}
\label{eq:Delta_Nd_squared_sqz_coh}
\Delta^2\hat{N}_d
=\cos^2\varphi\left(2\sinh^2r\cosh^2r+\vert\alpha\vert^2\right)
\nonumber\\
+\sin^2\varphi\left(\sinh^2r+\vert\alpha\vert^2e^{-r}\right)
\nonumber\\
+\sin^2\varphi\vert\alpha\vert^2\sinh2r\left(1
-\cos\left(2\theta_\alpha\right)\right)
\end{eqnarray}
where $\theta_\alpha$ is the phase of the coherent light. Squeezed states are typically characterized by $\xi=re^{i\theta}$ \cite{GerryKnight}, however, for simplicity, we considered $\xi=r\in\mathbb{R}^+$, therefore the phase of the squeezed vacuum was taken to be $\theta=0$. The minimum variance is obtained if $\theta_\alpha=0$ and using again Eq.~\eqref{eq:Delta_varphi_DEFINITION} we arrive at the phase sensitivity given by Eq.~\eqref{eq:Delta_phi_d_coh_sqz_vac_diff}. For a single detection scheme we have
\begin{equation}
\label{eq:N4_prime_sqz_coh}
\bigg\vert\frac{\partial\langle\hat{N}_4\rangle}{\partial\varphi}\bigg\vert=\frac{\vert\sin\varphi\vert}{2}\Big\vert \vert\alpha\vert^2-\sinh^2r\Big\vert
\end{equation}
and a variance
\begin{eqnarray}
\label{eq:Delta_N4_FINAL_sqz_coh}
\Delta^2\hat{N}_4
=\sin^4\left(\frac{\varphi}{2}\right)\frac{\sinh^22r}{2}
+\cos^4\left(\frac{\varphi}{2}\right)\vert\alpha\vert^2
\nonumber\\
+\frac{\sin^2\varphi}{4}\left(\sinh^2r+\frac{\vert\alpha\vert^2}{e^{2r}}
\right)
\nonumber\\
+\frac{\sin^2\varphi}{4}\sinh{2r}\vert\alpha\vert^2\left(1-\cos2\theta_\alpha\right)
\end{eqnarray}
Setting again $\theta_\alpha=0$ and employing the sensitivity definition from Eq.~\eqref{eq:Delta_varphi_DEFINITION} takes us to the result \eqref{eq:Delta_phi_sqz_coh_SINGLE_DET_theta_zero}.

\onecolumngrid

\bibliography{VBir_laser_based_bibliography}


\begin{table}
\centering
\renewcommand{\arraystretch}{1.2}
\begin{tabular}{c|c|c|c|c|c|c|c|}
\cline{2-8}
& \multicolumn{2}{c|}{\color{darksienna}Pump laser parameters} & {\color{darkblue}QED predicted} &  \multicolumn{2}{c|}{Achievable phase sensitivity} & \multicolumn{2}{c|}{Achievable phase sensitivity}\\
\arrayrulecolor{gray(x11gray)}\cline{2-3}
\arrayrulecolor{black}
& Maximum & Pulse & {\color{darkblue}phase shift} &  \multicolumn{2}{c}{$\boldsymbol{\Delta\varphi_{SQL}}$ given by Eq.~\eqref{eq:Delta_phi_SQL}} & \multicolumn{2}{|c|}{$\boldsymbol{\Delta\varphi_{SQL}}$ for a}\\
 & E-field & duration & $\boldsymbol{\Delta\varphi_{QED}}$   & \multicolumn{2}{c|}{for a CW probe laser of} &  \multicolumn{2}{c|}{pulsed probe laser of}\\
\arrayrulecolor{gray(x11gray)}\cline{5-8}
\arrayrulecolor{black}
  & $E_L$ [V/m] & $\tau_L$ [fs] & from Eq.~\eqref{eq:delta_phi_para_perp} & \mbox{ } $P=100$ W \mbox{ }& $P=500$ W &\mbox{ } $P=10^{10}$ W \mbox{ }& $P=10^{12}$ W\\
\arrayrulecolor{black}\hline
\hline
\multicolumn{1}{|c|}{ELI-NP\mbox{ }} & $10^{15}$ & $22$ & $\{6;10\}\times10^{-7}$   & $4\times10^{-4}$ & $1.8\times10^{-4}$  & $\boldsymbol{4.1\times10^{-8}}$   & $\boldsymbol{4.1\times10^{-9}}$ \\
\hline
\multicolumn{1}{|c|}{ELI-BL\mbox{ }} & $10^{15}$ & $150$ & $\{6;10\}\times10^{-7}$   & $1.5\times10^{-4}$ & $7\times10^{-5}$  & $\boldsymbol{1.5\times10^{-8}}$   & $\boldsymbol{1.5\times10^{-9}}$ \\
\hline
\multicolumn{1}{|c|}{Vulcan\mbox{ }} & $10^{14}$ & $500$ & $\{6;10\}\times10^{-9}$   & $8.6\times10^{-5}$ & $3.8\times10^{-5}$  & $8.6\times10^{-9}$   & $\boldsymbol{8.6\times10^{-10}}$ \\
\hline
\multicolumn{1}{|c|}{LFEX\mbox{ }} & $10^{14}$ & $10^4$ & $\{6;10\}\times10^{-9}$   & $1.9\times10^{-5}$ & $8.6\times10^{-6}$  & $\boldsymbol{1.9\times10^{-9}}$   & $\boldsymbol{1.9\times10^{-10}}$\\
\hline
\end{tabular}
\caption{\label{tab:Laser_parameters_and_sensitivities_COH}Comparison of QED-induced phase shifts $\boldsymbol{\Delta\varphi_{QED}}$ versus shot-noise (or SQL) limited phase sensitivities $\boldsymbol{\Delta\varphi_{SQL}}$ for four laser facilities: Extreme Light Infrastructure - Nuclear Physics (ELI-NP) $10$ PW laser, Extreme Light Infrastructure - Beamlines (ELI-BL) $10$ PW laser, Vulcan and LFEX (both $1$ PW lasers). The probe beam wavelength was taken $\lambda_p=532$ nm. Bold numerical values denote measurement sensitivities within the QED-predicted vacuum birefringence effect.}
\end{table}

\begin{table}
\centering
\renewcommand{\arraystretch}{1.2}
\begin{tabular}{c|c|c|c|c|c|c|c|}
\cline{2-8}
& \multicolumn{2}{c|}{\color{darksienna}Pump laser parameters} & {\color{darkblue}QED predicted} &  \multicolumn{2}{c|}{Theoretical phase sensitivity} & \multicolumn{2}{c|}{Phase sensitivity $\boldsymbol{\Delta\varphi_{CSV}}$ given}\\
\arrayrulecolor{gray(x11gray)}\cline{2-3}
\arrayrulecolor{black}
& Maximum & Pulse & {\color{darkblue}phase shift} &  \multicolumn{2}{c}{$\boldsymbol{\Delta\varphi_{HL}}$ given by Eq.~\eqref{eq:Delta_phi_Heisenberg_limit} for} & \multicolumn{2}{|c|}{by Eq.~\eqref{eq:Delta_phi_coherent_squeezed_approx} for a CW probe with}\\
 & E-field & duration & $\boldsymbol{\Delta\varphi_{QED}}$   & \multicolumn{2}{c|}{a CW probe laser power of} &  \multicolumn{2}{c|}{$P=200$ W and squeezing factor}\\
\arrayrulecolor{gray(x11gray)}\cline{5-8}
\arrayrulecolor{black}
  & $E_L$ [V/m] & $\tau_L$ [fs] & from Eq.~\eqref{eq:delta_phi_para_perp} & \mbox{ } $P=10$ W \mbox{ }& $P=100$ W &\mbox{  } $r=3.5$ \mbox{  } & $r=6$ \\
\arrayrulecolor{black}\hline
\hline
\multicolumn{1}{|c|}{ELI-NP\mbox{ }} & $10^{15}$ & $22$ & $\{6;10\}\times10^{-7}$  & $1.6\times10^{-6}$   & $\boldsymbol{1.6\times10^{-7}}$ & $8.7\times10^{-6}$ & $\boldsymbol{7\times10^{-7}}$\\
\hline
\multicolumn{1}{|c|}{ELI-BL\mbox{ }} & $10^{15}$ & $150$ & $\{6;10\}\times10^{-7}$    & $\boldsymbol{2.5\times10^{-7}}$   & $\boldsymbol{2.5\times10^{-8}}$ & $3.3\times10^{-6}$ & $\boldsymbol{2.7\times10^{-7}}$\\
\hline
\multicolumn{1}{|c|}{Vulcan\mbox{ }} & $10^{14}$ & $500$ & $\{6;10\}\times10^{-9}$  & $7\times10^{-8}$   & $\boldsymbol{7\times10^{-9}}$ & $1.8\times10^{-6}$ & $1.5\times10^{-7}$\\
\hline
\multicolumn{1}{|c|}{LFEX\mbox{ }} & $10^{14}$ & $10^4$ & $\{6;10\}\times10^{-9}$     & $\boldsymbol{3.7\times10^{-9}}$   & $\boldsymbol{3.7\times10^{-10}}$ & $4.7\times10^{-7}$ & $3\times10^{-8}$\\
\hline
\end{tabular}
\caption{\label{tab:Laser_parameters_and_sensitivities_SQZ}Comparison of QED-induced phase shifts $\boldsymbol{\Delta\varphi_{QED}}$ versus theoretical best achievable sensitivities using non-classical light $\boldsymbol{\Delta\varphi_{HL}}$, as well as sub-optimal but more realistic sensitivities for the coherent-squeezed vacuum scenario, $\boldsymbol{\Delta\varphi_{CSV}}$. Bold numerical values denote measurement sensitivities within the QED-predicted vacuum birefringence effect.}
\end{table}

\twocolumngrid

\end{document}